\documentclass{article}

\usepackage{amssymb,amsfonts,amsmath}
\usepackage{cite,enumerate,float,indentfirst}
\usepackage{color}

\def\be{\begin{eqnarray}}
\def\ee{\end{eqnarray}}
\def\nn{\nonumber}

\hoffset= - 1.0in         

\begin{document}

\hfill ITEP/TH-33/17

\hfill IITP/TH-20/17

\bigskip

\centerline{\Large{
HOMFLY for twist knots and exclusive Racah matrices in representation $[333]$
}}

\bigskip

\bigskip

\centerline{\bf  A.Morozov }

\bigskip

{\footnotesize
\centerline{{\it
ITEP, Moscow 117218, Russia}}

\centerline{{\it
Institute for Information Transmission Problems, Moscow 127994, Russia
}}

\centerline{{\it
National Research Nuclear University MEPhI, Moscow 115409, Russia
}}
}

\bigskip

\bigskip

\centerline{ABSTRACT}

\bigskip

{\footnotesize
Next step is reported in the program of Racah matrices extraction from the
differential expansion of HOMFLY polynomials for twist knots:
from the double-column rectangular representations $R=[rr]$ to a triple-column
and triple-hook $R=[333]$.
The main new phenomenon is the deviation of the particular coefficient
$f_{[332]}^{[21]}$ from the corresponding skew dimension,
what opens a way to further generalizations.
}

\bigskip


\section{Introduction}

Calculation of Racah matrices is the long-standing, difficult and challenging
problem in theoretical physics \cite{racahmatrices}.
It is further obscured by the basis-dependence of the answer in the case of generic
representations, but this "multiplicity problem" is absent in the case of
rectangular representations.
The modern way \cite{M3141}-\cite{MnonrectS}
to evaluate the most important "exclusive" matrices $\bar S_{\mu\nu}^{R}$,
$$
\Big((R\otimes \bar R)\otimes R \longrightarrow R \Big)
\ \stackrel{\bar S}{\longrightarrow} \
\Big(R\otimes (\bar R \otimes R) \longrightarrow R \Big)
$$
is based on the combination of two very different expressions for
$R$-colored HOMFLY polynomials \cite{knotpols} of the double-braid knots,
coming one from the arborescent calculus of \cite{arbor}
and another from the differential expansion theory \cite{IMMMfe,evo,diffexpan}
in the case of rectangular $R=[r^s]$ with $s$ columns of length $r$:
\be
\boxed{
H_R^{(m,n)}\!\!
= \!\! \sum_{\mu,\nu\subset R}
\frac{\sqrt{{\cal D}_\mu{\cal D}_\nu}}{d_R}\,\bar S_{\mu\nu}^{R}\,
\Lambda_\mu^m\Lambda_\nu^n
=  \sum_{\lambda\subset R} \chi^*_{\lambda^{tr}}(r)\chi^*_\lambda(s)\cdot
\{q\}^{2|\lambda|} \cdot h_\lambda^2\cdot {\chi^*_\lambda(N+r)\chi^*_\lambda(N-s)}
\cdot
\sum_{\mu,\nu\in\lambda}
\frac{f_\lambda^\mu f_\lambda^\nu}{F_\lambda^{(1)}} \cdot\Lambda_\mu^m\Lambda_\nu^n
}
\label{HRdb1}
\ee
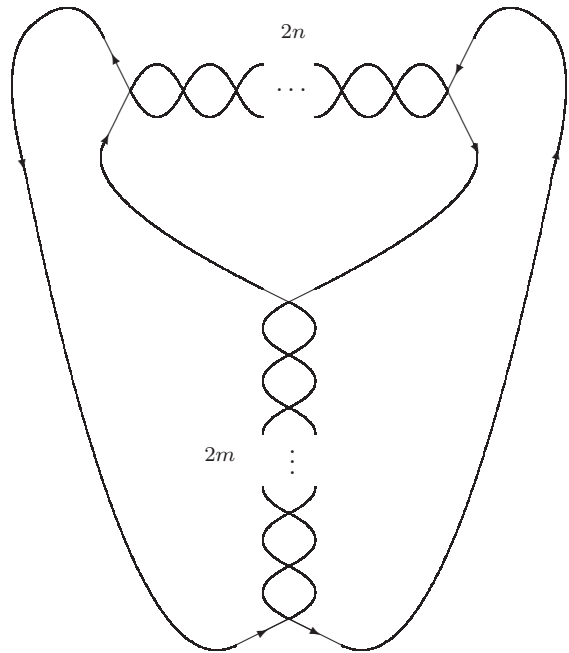
\begin{picture}(200,280)(-230,-230)
\qbezier(-40,0)(-50,20)(-60,0)
\qbezier(-40,0)(-50,-20)(-60,0)
\qbezier(-20,0)(-30,20)(-40,0)
\qbezier(-20,0)(-30,-20)(-40,0)
\qbezier(-20,0)(-15,10)(-10,10)
\qbezier(-20,0)(-15,-10)(-10,-10)
\put(-5,0){\mbox{$\ldots$}}
\qbezier(10,10)(15,10)(20,0)
\qbezier(10,-10)(15,-10)(20,0)
\qbezier(20,0)(30,20)(40,0)
\qbezier(20,0)(30,-20)(40,0)
\qbezier(40,0)(50,20)(60,0)
\qbezier(40,0)(50,-20)(60,0)
\put(-60,0){\line(-1,2){10}}
\put(-60,0){\line(-1,-2){10}}
\put(60,0){\line(1,2){10}}
\put(60,0){\line(1,-2){10}}
\qbezier(0,-80)(-20,-90)(0,-100)
\qbezier(0,-80)(20,-90)(0,-100)
\qbezier(0,-100)(-20,-110)(0,-120)
\qbezier(0,-100)(20,-110)(0,-120)
\qbezier(0,-120)(-10,-125)(-10,-130)
\qbezier(0,-120)(10,-125)(10,-130)
\put(0,-145){\mbox{$\vdots$}}
\qbezier(0,-160)(-10,-155)(-10,-150)
\qbezier(0,-160)(10,-155)(10,-150)
\qbezier(0,-160)(-20,-170)(0,-180)
\qbezier(0,-160)(20,-170)(0,-180)
\qbezier(0,-180)(-20,-190)(0,-200)
\qbezier(0,-180)(20,-190)(0,-200)
\put(0,-80){\line(-2,1){10}}
\put(0,-80){\line(2,1){10}}
\put(0,-200){\line(-2,-1){10}}
\put(0,-200){\line(2,-1){10}}
\put(0,-200){\line(-2,-1){20}}
\put(0,-200){\line(2,-1){20}}
\qbezier(-10,-75)(-80,-40)(-70,-20)
\qbezier(10,-75)(80,-40)(70,-20)
\put(-10,-205){\vector(2,1){2}}
\put(10,-205){\vector(2,-1){2}}
\put(-65,10){\vector(-1,2){2}}
\put(65,10){\vector(-1,-2){2}}
\put(-70,-20){\vector(1,2){2}}
\put(70,-20){\vector(1,-2){2}}
\put(-3,20){\mbox{\footnotesize$2n$}}
\put(-32,-140){\mbox{\footnotesize $2m$}}
%
\qbezier(-70,20)(-80,40)(-97,25)
\qbezier(-97,25)(-111,13))(-100,-30)
\qbezier(-100,-30)(-60,-230)(-20,-210)
\qbezier(70,20)(80,40)(97,25)
\qbezier(97,25)(111,13))(100,-30)
\qbezier(100,-30)(60,-230)(20,-210)
\put(-102,-22){\vector(1,-4){2}}
\put(100,-30){\vector(1,4){2}}
\end{picture}

\noindent
Here the sums go over sub-diagrams of the Young diagram $R$,
and $\chi^*(N)$ denote the corresponding dimensions for the algebra $sl_N$,
i.e. the values of Schur functions $\chi\{p_k\}$ at the topological locus
$p_k=p_k^*=\frac{\{A^k\}}{\{q^k\}}$ with $\{x\} = x-x^{-1}$ and $A=q^N$.
Combinatorial factor $h_\lambda^2$ cancels the $N$-independent denominators in
$\chi^*_\lambda(N+r)\chi^*_\lambda(N-s)$, converting it into a product of "differentials" $\{Aq^i\}$.
The other ingredients of the formula come from the evolution method \cite{DMMSS,evo}
applied to the family of twist knots (double braids with $n=1$):
as functions of the "evolution parameter" $m$ knot polynomials are then decomposed
into sums of representation $\mu \in R\otimes\bar R$
(which for rectangular $R$ can be labeled by sub-diagrams of the $R$ itself)
with dimensions ${\cal D}_\mu$,
and $m$-dependence is then provided by $m$-th power of the "eigenvalue" $\Lambda_\mu$
of the ${\cal R}$-matrix,
a $q$-power of the Casimir or cut-and-join operator \cite{DMMSS}.
In arborescent calculus the weights are made from the elements of Racah matrix $\bar S$
while in the theory of differential expansions they are composed into amusing
generating functions \cite{evo}
\be
F_\lambda^{(m)}(q,A) =
\frac{c_\lambda}{\prod_{(\alpha,\beta)\in \lambda} \{A\,q^{\alpha-\beta}\}}
\sum_{\mu\subset\lambda} f_\lambda^\mu(q,A)\cdot \Lambda_\mu^m
\ee
with $f_\lambda^\emptyset=1$.
Each term in the sum has a non-trivial denominator, however the full sum is
a Laurent polynomial in $A$ and $q$ for all $m$.
Moreover, it vanishes for $m=0$ (unknot), equals one for $m=-1$ (figure eight knot $4_1$)
and is a monomial at $m=1$ (trefoil).
According to \cite{Mfact} and \cite{KMtwist} the $F$-functions are best described
in a peculiar hook parametrization of Young diagrams:

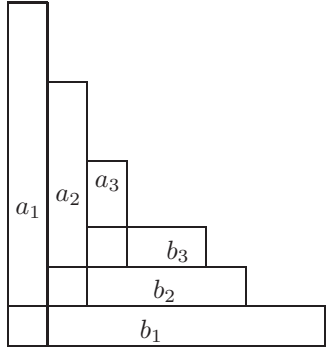
\begin{picture}(300,150)(-50,-10)
\put(0,0){\line(0,1){130}}
\put(15,0){\line(0,1){130}}
\put(30,15){\line(0,1){85}}
\put(45,30){\line(0,1){40}}
\put(75,30){\line(0,1){15}}
\put(90,15){\line(0,1){15}}
\put(120,0){\line(0,1){15}}

\put(0,0){\line(1,0){120}}
\put(0,15){\line(1,0){120}}
\put(15,30){\line(1,0){75}}
\put(30,45){\line(1,0){45}}
\put(0,130){\line(1,0){15}}
\put(15,100){\line(1,0){15}}
\put(30,70){\line(1,0){15}}

\put(3,50){\mbox{$a_1$}}
\put(18,55){\mbox{$a_2$}}
\put(33,60){\mbox{$a_3$}}
\put(50,3){\mbox{$b_1$}}
\put(55,18){\mbox{$b_2$}}
\put(60,33){\mbox{$b_3$}}

\put(130,100){\mbox{a 3-hook Young diagram $\ (a_1,b_1|a_2,b_2|a_3,b_3)$}}
\put(130,80){\mbox{$= \, [a_1+1,a_2+2,a_3+3,3^{b_3},2^{b_2-b_3-1},1^{b_1-b_2-1}]$}}
\put(130,60){\mbox{of the size $\ a_1+a_2+a_3+b_1+b_2+b_3+3$}}

\end{picture}

\noindent
In particular,
\vspace{-0.2cm}
\be
\Lambda_\mu = \Lambda_{(i_1,j_1|i_2,j_2|\ldots)} =
\prod_{k=1}  (A\cdot q^{i_k-j_k})^{2(i_k+j_k+1)}
\ee
the overall coefficients
\vspace{-0.2cm}
\be
c_\lambda = c_{(a_1,b_1|a_2,b_2|\ldots)} =
\prod_{k=1}  (A\cdot q^{\frac{a_k-b_k}{2}})^{(a_k+b_k+1)}
\ee
and
\be
F_\lambda^{(-1)}=1, \ \ \ \ \ F_\lambda^{(0)}=\delta_{\lambda,\emptyset}, \ \ \ \ \
F_\lambda^{(1)} = (-)^{\sum_{k}  (a_k+b_k+1)}c_\lambda^2
\label{sumrules1}
\ee
Clearly, $c_\lambda$ drops away from the r.h.s. of (\ref{HRdb1}).

The shape of the coefficients $f_\lambda^\mu$ strongly depends on the number of hooks
in $\lambda$ and $\mu$.
Currently they are fully known for $\lambda=(a_1,b_1|a_2,0)$ --
what is enough to get the Racah matrices $\bar S$ for the case $R=[r,r]$
(actually, for this purpose $b_1=0,1$ is sufficient).
understand.

$\bullet$ As already mentioned, for the empty diagram $\mu$ always
\vspace{-0.3cm}
\be
f_\lambda^\emptyset = 1
\ee

$\bullet$
For the single-hook $\lambda$ and thus single-hook $\mu\subset\lambda$
expressions are still relatively simple and fully factorized:
\be
f_{(a,b)}^{(i,j)} = g_{(a,b)}^{(i,j)} \cdot K_{(a,b)}^{(i,j)}
= (-)^{i+j+1}
\cdot \frac{[a]!}{[a-i]![i]!}\cdot\frac{[b]!}{[b-j]![j]!}\cdot\frac{[a+b+1]}{[i+j+1]}
\cdot \frac{D_a!D_i!}{D_{a+i+1}!}\cdot\frac{\bar D_b!\bar D_j!}{\bar D_{b+j+1}!}
\cdot \frac{D_{2i+1}D_{-2j-1}}{D_0D_{i-j}}
\label{f1vsgK1}
\ee
with
\vspace{-0.5cm}
\be
g_{(a,b)}^{(i,j)}
= (-)^{i+j+1} \cdot \frac{D_{2i+1}\bar D_{2j+1}}{D_0D_{i-j}}\cdot
\frac{(D_a!)^2}{D_{a+i+1}!\,D_{a-i-1}!}\cdot\frac{(\bar D_b!)^2}{\bar D_{b+j+1}!\,\bar
D_{b-j-1}!}
\label{gfactors1}
\ee
and
\vspace{-0.5cm}
\be
K_{\lambda}^{\mu}(N) = \frac{\chi^*_{\lambda/\mu}(N)\,\chi^*_{\mu}(N)}{\chi^*_{\lambda}(N)}
\ee
This $K$ involves skew characters, defined by
\be
\sum_{\mu\subset \lambda} \chi_{\lambda/\mu}\{p'_k\}\cdot \chi_{\mu^{tr}}\{p''_k\} =
\chi_\lambda\{p'_k+p''_k\}
\label{skewdef}
\ee
and satisfying the sum rule
\vspace{-0.4cm}
\be
\sum_{\mu\subset \lambda} (-)^{|\mu|}\cdot\chi_{\lambda/\mu}\cdot \chi_{\mu^{tr}}
= \delta_{\lambda,\emptyset}
\label{naivesumrule1}
\ee
which follows from (\ref{skewdef}) and the transposition law
$\chi_\mu\{-p_k\} = (-)^{|\mu|} \chi_{\mu^{tr}}\{p_k\}$,
and can be considered as a prototype of (\ref{sumrules1}).
The other notation in (\ref{f1vsgK1}) and (\ref{gfactors1}) is:
\be
D_a=\{Aq^a\}={\{q\}}\cdot [N+a], \ \ \ \ \ \bar D_b=\{A/q^b\}={\{q\}}\cdot [N-b]
\ee
and
\be D_a!=\prod_{k=0}^a D_k
= {\{q\}}^{a+1}\cdot \frac{[N+a]!}{[N-1]!},  \ \ \ \ \ \ \ \
\bar D_b! = \prod_{k=0}^b \bar D_k =  {\{q\}}^{b+1}\cdot\frac{[N]!}{[N-b-1]!}
\ee
(note that these products start from $k=0$ and include
respectively $a+1$ and $b+1$ factors).

$\bullet$ For two-hook $\lambda=(a_1,b_1|a_2,b_2)$ the formulas are far more involved,
and they are different for different number of hooks in $\mu$:
\be
f_{(a_1,b_1|a_2,b_2)}^{(i_1,j_1)}
= f^{(a_1,b_1)}_{(i_1,j_1)} \cdot \xi_{(a_1,b_1|a_2,b_2)}^{(i_1,j_1)}
= g_{(a_1,b_1)}^{(i_1,j_1)}\cdot K_{(a_1,b_1)}^{(i_1,j_1)}(\underline{N})
\cdot \xi_{(a_1,b_1|a_2,b_2)}^{(i_1,j_1)}
\label{f12}
\ee
\be
f_{(a_1,b_1|a_2,b_2)}^{(i_1,j_1|i_2,j_2)} =
\frac{[N+i_1+i_2+1][N-j_1-j_2-1]}{[N+i_1-j_2][N+i_2-j_1]}
\cdot \underbrace{g_{(a_1,b_1)}^{(i_1,j_1)}\,g_{(a_2,b_2)}^{(i_2,j_2)}
\cdot K_{(a_1,b_1 )}^{(i_1,j_1 )}(\underline{ N})
K_{( a_2,b_2)}^{(
i_2,j_2)}(\underline{N})}_{f_{(a_1,b_1)}^{(i_1,j_1)}\,f_{(a_2,b_2)}^{(i_2,j_2)}}
\cdot\, \xi_{(a_1,b_1|a_2,b_2)}^{(i_1,j_1|i_2,j_2)}
\ee
Non-trivial are the correction factors,  $\ {\bf true \ for \ a_2\cdot b_2=0}$:
\be
\xi_{(a_1,b_1|a_2,b_2)}^{(i_1,j_1)} =
\frac{[N+a_2-j_1][N-b_2+i_1]}{[N+a_2+i_1+1][N-b_2-j_1-1]}
\cdot
{\cal K}_{(a_1,b_1|a_2,b_2)}^{(i_1,j_1)}(\underline{N+i_1-j_1})
\cdot \boxed{\delta_{i_1\cdot j_1 }}\ +
\nn\ee
\vspace{-0.4cm}
\be
\!\!\!\!\!
+ \
{\cal K}_{(a_1,b_1|a_2,b_2)}^{(i_1,j_1)}
 \Big(\underline{N+(i_1+1)\delta_{b_2 }-(j_1+1)\delta_{a_2 }}\Big)
 \cdot\ \boxed{(1-\delta_{i_1\cdot j_1 })}
\label{xi1}
\ee
and
\be
\!\!\!
\xi_{(a_1,b_1|a_2,b_2)}^{(i_1,j_1|i_2,j_2)} =
 {\cal K}_{(a_1,b_1|a_2,b_2)}^{(i_1,j_1|i_2,j_2)}
\Big(\underline{N+(i_1+i_2+2)\cdot \delta_{b_2 } - (j_1+j_2+2)\cdot \delta_{a_2 }}\Big)
\label{xi2}
\ee
where
$\ \delta_x = \left\{\begin{array}{ccc}  1 & {\rm for} & x = 0 \\ 0 & {\rm for} & x\neq 1
\end{array}\right. \ $
and
\be
{\cal K}_{(a_1,b_1|a_2,b_2)}^{(i_1,j_1)}(\underline{N}) =
\frac{K_{(a_1,b_1|a_2,b_2)}^{(i_1,j_1)}(\underline{N})}
{K_{(a_1,b_1 )}^{(i_1,j_1)}(\underline{N})}
\ee
\be
{\cal K}_{(a_1,b_1|a_2,b_2)}^{(i_1,j_1|i_2,j_2)}(\underline{N}) =
\frac{K_{(a_1,b_1|a_2,b_2)}^{(i_1,j_1|i_2,j_2)}(\underline{N})}
{K_{(a_1,b_1 )}^{(i_1,j_1)}(\underline{N})\cdot K_{(a_2,b_2 )}^{(i_2,j_2)}(\underline{N})
}
\label{calK22}
\ee
Thus corrections involve a natural modification of $K$-factors and
somewhat strange shifts of the argument $N$, i.e. multiplicative shift of $A$
by powers of $q$.
These formulas were found in \cite{Mfact,KMtwist} for the case when $a_2\cdot b_2=0$
(i.e. when either $b_2=0$ or $a_2=0$).
Sufficient for all the simplest non-symmetric rectangular representations
$R=[r,r]$ and $R=[2^r]$ are respectively $b_2=0$ and $a_2=0$.
Note that underlined expression are
the {\it arguments} of ${\cal K}$-functions --
{\it not} additional algebraic factors.
Boxes contain projectors on sectors with particular values of $i_1$ and $j_1$.

Our goal in this paper is to make the first step towards lifting the
restriction $a_2\cdot b_2=0$.
Namely, we consider the case of the simplest 3-hook $R=[333]$,
which has $20$ Young sub-diagrams, of which there are two,
$\lambda = [332]=(2,2|1,1)$ and $\lambda=[333]=(2,2|1,1|0,0)$
with $a_2\cdot b_2\neq 0$.

\section{The new function $F_{(22|11)}^{(m)}=F_{[332]}^{(m)}$}

The diagram $[332]=(22|11)$ is still two-hook, but both $a_2=b_2=1$ are non-vanishing.
If we  apply just the same formulas (\ref{f12})-(\ref{calK22}) in this case,
the answer will be non-polynomial.
However, one can introduce additional correction factors $\eta_\lambda^\mu$ for all the
items
in the sum over $\mu$ and adjust them to cancel all the singularities.
Of $19$ factors non-trivial (different from unity)   are just $8$
(we omit the subscript $\lambda=(22|11)$ to simplify the formulas):
\be
\eta^\emptyset = \eta^{(00)}=\eta^{(01)}
=\eta^{(10)}=\eta^{(02)}=\eta^{(20)}=\eta^{(22)}=1
\nn \\ \nn \\
\eta^{(11)}
= \frac{\frac{[4]^3}{[5][2]}\frac{D_1D_{-1}}{D_2D_{-2}} }{{\cal
K}_{(22|11)}^{(11)}(\underline{N})},
\ \ \ \ \ \ \ \ \
\eta^{(12)} = \frac{{\cal K}_{(22|11)}^{(12)}(\underline{N+2})}
{{\cal K}_{(22|11)}^{(12)}(\underline{N})},
\ \ \ \ \
\eta^{(21)} = \frac{{\cal K}_{(22|11)}^{(21)}(\underline{N-2})}
{{\cal K}_{(22|11)}^{(21)}(\underline{N})}\nn \\
\eta^{(11|00)} = \frac{D_2D_0^2D_{-2}}{D_3D_1D_{-1}D_{-3}}, \ \ \ \ \ \ \ \ \
\eta^{(12|00)} = \eta^{(12|01)}=\frac{D_2D_0}{D_3D_{-1}}, \ \ \ \ \ \ \
\eta^{(21|00)}=\eta^{(21|10)}=\frac{D_0D_{-2}}{D_3D_{-1}} \nn \\ \nn \\ \nn\\
\eta^{(22|00)}=\eta^{(22|01)}=\eta^{(22|10)}
=\eta^{(22|11)}=1
\ee
and the resulting expression is
\be
A^{-8}F_{(22|11)}^{(m)} =
\frac{1}{D_{2}D_1^2D_0^2D_{-1}^2D_{-2}}
-\frac{[4][2]}{D_3D_2D_1D_0^2D_{-1}D_{-2}D_{-3}}\cdot \Lambda_{(00)}^m + \nn \\
+ \frac{\frac{[4]}{[2]}\cdot
\Big(A\,(q^4+2q^2+2+q^{-2}+q^{-4})-A^{-1}\,(q^4+q^2+2+2q^{-2}+q^{-4})\Big)}
{D_3D_2^2D_1D_0D_{-1}^2D_{-2}D_{-4}}\cdot \Lambda_{(01)}^m
+ \nn \\
+\frac{\frac{[4]}{[2]}\cdot
\Big(A\,(q^4+q^2+2+2q^{-2}+q^{-4})-A^{-1}\,(q^4+2q^2+2+q^{-2}+q^{-4})\Big)}
{D_4D_2D_1^2D_0D_{-1}D_{-2}^2D_{-3}} \cdot \Lambda_{(10)}^m
- \nn \\
-\frac{\frac{[5][4]}{[3]}}{D_3D_2^2D_1D_0D_{-2}D_{-3}D_{-4}}\cdot \Lambda_{(02)}^m
-\frac{\frac{[5][4]}{[3]}}{D_4D_3D_2 D_0D_{-1}D_{-2}^2D_{-3}}\cdot \Lambda_{(20)}^m
- \frac{\frac{[4]^3[2]}{[3]} }
{D_4D_2^2D_0^2D_{-2}^2D_{-4}}\cdot \Lambda_{(11)}^m
+ \nn \\
+ \frac{[5]\cdot\Big(A \,(q^4+q^2+2 +q^{-2})-A^{-1}\,(q^2+2 +q^{-2}+q^{-4})\Big)}
{D_4D_2^2D_1D_0D_{-1}^2D_{-3}D_{-4}}\cdot \Lambda_{(12)}^m
+\nn \\
+ \frac{[5]\cdot\Big(A\,(q^2+2+q^{-2}+q^{-4})-A^{-1}\,(q^4+q^2+2+q^{-2} )\Big)}
{D_4D_3D_1^2D_0D_{-1}D_{-2}^2D_{-4}}\cdot \Lambda_{(21)}^m
- \nn \\
- \frac{[4]^2}{D_4D_3D_1D_0^2D_{-1}D_{-3}D_{-4}}\cdot \Lambda_{(22)}^m
+ \frac{\frac{[5][4]^2}{[2]^2}}{D_4D_3D_1D_0^2D_{-1}D_{-3}D_{-4}} \cdot \Lambda_{(11|00)}^m
- \nn \\
-\frac{[5][4]}{D_4D_3D_1D_0D_{-1}D_{-2}^2D_{-4}}\cdot \Lambda_{(12|00)}^m
-\frac{[5][4]}{D_4D_2^2D_1D_0D_{-1}D_{-3}D_{-4}}\cdot \Lambda_{(21|00)}^m
+ \nn \\
+ \frac{\frac{[5][4]}{[3][2]}}{D_4D_3D_0D_{-1}^2D_{-2}^2D_{-3}}\cdot \Lambda_{(12|01)}^m
+ \frac{\frac{[5][4]}{[3][2]}}{D_3D_2^2D_1^2D_0D_{-3}D_{-4}}\cdot \Lambda_{(21|10)}^m
+ \frac{\frac{[4]^2[2]^2}{[3]}}{D_4D_2^2D_0^2D_{-2}^2D_{-4}}\cdot \Lambda_{(22|00)}^m
- \nn \\
- \frac{[4]}{D_4D_2D_1D_0D_{-1}D_{-2}^2D_{-3}}\cdot \Lambda_{(22|01)}^m
- \frac{[4]}{D_3D_2^2D_1D_0D_{-1}D_{-2}D_{-4}}\cdot \Lambda_{(22|10)}^m
+ \nn \\
+  \frac{1}{D_3D_2D_1D_0^2D_{-1}D_{-2}D_{-3}}\cdot \Lambda_{(22|11)}^m
\ee
It nicely satisfies the sum rules (\ref{sumrules1}).


\section{Extension to $F_{(a_1b_1|11)}$}

We can now develop the success with $F_{(22|11)}$ and extend it to other
2-hook diagrams with $a_2\cdot b_2\neq 0$.
We actually restrict our attention to the case of $a_2\cdot b_2=1$, i.e.
$a_2=b_2=1$.

In the next case of $F_{(33|11)}$ the correction factors are
(again we write just $\eta^\mu$ instead of $\eta_{(33|11)}^\mu$):
\be
\eta^\emptyset = \eta^{(00)}=\eta^{(01)}
=\eta^{(10)}=\eta^{(02)}=\eta^{(20)}=
\eta^{(30)} = \eta^{(03)} = \eta^{(22)} = \eta^{(32)}=\eta^{(23)} = \eta^{(33)} =1
\nn \\ \nn \\
\eta^{(11)}
= \frac{u_{(33)} }{{\cal K}_{(33|11)}^{(11)}(\underline{N})},
\ \ \ \ \ \ \ \ \
\eta^{(12)} = \frac{{\cal K}_{(33|11)}^{(12)}(\underline{N+2})}
{{\cal K}_{(33|11)}^{(12)}(\underline{N})}
\ \ \ \ \
\eta^{(21)} = \frac{{\cal K}_{(33|11)}^{(21)}(\underline{N-2})}
{{\cal K}_{(33|11)}^{(21)}(\underline{N})}\nn \\
\eta^{(13)} = \frac{{\cal K}_{(33|11)}^{(13)}(\underline{N+2})}
{{\cal K}_{(33|11)}^{(13)}(\underline{N})}
\ \ \ \ \
\eta^{(31)} = \frac{{\cal K}_{(33|11)}^{(31)}(\underline{N-2})}
{{\cal K}_{(33|11)}^{(31)}(\underline{N})}\nn \\
\eta^{(11|00)} = \frac{D_2D_0^2D_{-2}}{D_3D_1D_{-1}D_{-3}} \ \ \ \ \ \ \ \ \ \nn \\
\eta^{(12|00)} = \eta^{(12|01)}=\eta^{(13|00)} = \eta^{(13|01)}=\frac{D_2D_0}{D_3D_{-1}} \ \
\ \ \ \ \
\eta^{(21|00)}=\eta^{(21|10)}=\eta^{(31|00)}=\eta^{(31|10)}=\frac{D_0D_{-2}}{D_3D_{-1}}
\nn \\ \nn \\ \nn\\
\eta^{(22|00)}=\eta^{(22|01)}=\eta^{(22|10)}=\eta^{(22|11)} =
\eta^{(23|00)}=\eta^{(23|01)}=\eta^{(23|10)}=\eta^{(23|11)} = \nn \\
\eta^{(32|00)}=\eta^{(32|01)}=\eta^{(32|10)}=\eta^{(32|11)}=
\eta^{(33|00)}=\eta^{(33|01)}=\eta^{(33|10)}=\eta^{(33|11)}=1
\ee
This implies a simple extension of (\ref{xi1}) and (\ref{xi2}) to arbitrary diagrams
$(a_1,b_1|1,1)$, i.e.   ${\bf  true\ for \ a_2\cdot b_2=0,1}$ are:
\be
\xi_{(a_1,b_1|a_2,b_2)}^{(i_1,j_1)} =
\frac{[N+a_2-j_1][N-b_2+i_1]}{[N+a_2+i_1+1][N-b_2-j_1-1]}
\cdot
{\cal K}_{(a_1,b_1|a_2,b_2)}^{(i_1,j_1)}(\underline{N+i_1-j_1})
\cdot \boxed{\delta_{i_1\cdot j_1 }}\ +
\nn\ee
\vspace{-0.6cm}
\be
\!\!\!\!\!\!\!\!\!\!\!\!\!\!\!\!
+ \
{\cal K}_{(a_1,b_1|a_2,b_2)}^{(i_1,j_1)}
 \Big(\underline{N+(i_1+1)\delta_{b_2}-(j_1+1)\delta_{a_2}+2(1-\delta_{a_2\cdot b_2})
 \big(\delta_{i_1-1}-\delta_{j_1-1}\big)}\Big)
 \cdot\ \boxed{(1-\delta_{i_1\cdot j_1}) }
 \cdot\left(\frac{u_{(a_1,b_1|a_2,b_2)}}{{\cal K}_{(a_1,b_1|a_2,b_2)}^{(1|1)}}\right)
^{(1-\delta_{a_2\cdot b_2})\cdot \delta_{i_1\cdot j_1-1}}
\nn\label{xi1a}
\ee
with
\be
\boxed{
u_{(a_1,b_1|1,1)} = \left({\cal K}_{(a_1,b_1|1,1)}^{(1,1)}
- \frac{[a_1+2]\,[b_1+2]}{[a_1]\,[b_1]}
\cdot\frac{[3][2]^2\{q\}^2}{D_0^2}\right)\cdot
\frac{D_1D_0^2D_{-1}}{D_3D_2D_{-2}D_{-3}}
}
\label{nonskew}
\ee
and
\be
\!\!\!\!\!\!\!\!\!\!\!
\xi_{(a_1,b_1|a_2,b_2)}^{(i_1,j_1|i_2,j_2)} =
{\cal K}_{(a_1,b_1|a_2,b_2)}^{(i_1,j_1|i_2,j_2)}
\Big(\underline{N+(i_1+i_2+2)\cdot \delta_{b_2} - (j_1+j_2+2)\cdot \delta_{a_2}}\Big)
\cdot
\left(\frac{D_2D_0}{D_3D_{-1}}\right)^{\delta_{i_1-1}\cdot(1-\delta_{a_1\cdot b_1})}
\!\!\cdot
\left(\frac{D_0D_{-2}}{D_1D_{-3}}\right)^{\delta_{j_1-1}\cdot(1-\delta_{a_1\cdot b_1})}
\nn\label{xi2a}
\ee

Formula (\ref{nonskew}) means that the coefficient $f^{(11)}_\lambda$  is no longer
proportional to the skew character $\chi^*_{\lambda/(11)}$.
Interpretation of this deviation remains to be found.
Note that for $a_2\cdot b_2=0$ we have just
\be
u_{(a_1,b_1|a_2,b_2)} =  {\cal K}_{(a_1,b_1|a_2,b_2)}^{(1,1)}\ \ \ \
{\rm for} \ a_2\cdot b_2=0
\ee
instead of (\ref{nonskew}) -- as one more manifestation of discontinuity
of the formulas, expressed in terms of hook variables.

\section{The new function $F_{[333]}^{(m)}=F_{(22|11|00)}^{(m)}$}

This $F$-factor is the first, associated with the triple-hook diagram $\lambda$.
To get an explicit formula we impose the polynomiality requirement
on the correction factors $\eta_{(22|11|00)}^\mu$
to the naive analogue of (\ref{f12})-(\ref{calK22})
for 3-hook diagrams:
\be
f_{(a_1,b_1|a_2,b_2|a_3,b_3)}^{(i_1,j_1)}
= f^{(a_1,b_1)}_{(i_1,j_1)} \cdot \xi_{(a_1,b_1|a_2,b_2)}^{(i_1,j_1)}
= g_{(a_1,b_1)}^{(i_1,j_1)}\cdot K_{(a_1,b_1)}^{(i_1,j_1)}(N)
\cdot \xi_{(a_1,b_1|a_2,b_2|a_3,b_3)}^{(i_1,j_1)}
\nn\ee
\be
f_{(a_1,b_1|a_2,b_2|a_3,b_3)}^{(i_1,j_1|i_2,j_2)} =
\frac{[N+i_1+i_2+1][N-j_1-j_2-1]}{[N+i_1-j_2][N+i_2-j_1]}
\cdot \underbrace{g_{(a_1,b_1)}^{(i_1,j_1)}\,g_{(a_2,b_2)}^{(i_2,j_2)}
\cdot K_{(a_1,b_1 )}^{(i_1,j_1 )}(N)
K_{( a_2,b_2)}^{( i_2,j_2)}(N)}_{f_{(a_1,b_1)}^{(i_1,j_1)}\,f_{(a_2,b_2)}^{(i_2,j_2)}}
\cdot\,\xi_{(a_1,b_1|a_2,b_2|a_3,b_3)}^{(i_1,j_1|i_2,j_2)}
\nn\ee
\be
\!\!\!\!\!\!\!\!\!\!
f_{(a_1,b_1|a_2,b_2|a_3,b_3)}^{(i_1,j_1|i_2,j_2|i_3,j_3)} =
\frac{[N+i_1+i_2+1][N-j_1-j_2-1]}{[N+i_1-j_2][N+i_2-j_1]}
\cdot
\frac{[N+i_1+i_3+1][N-j_1-j_3-1]}{[N+i_1-j_3][N+i_3-j_1]}
\cdot
\frac{[N+i_2+i_3+1][N-j_2-j_3-1]}{[N+i_2-j_3][N+i_3-j_2]}\cdot
\nn \ee
\vspace{-0.4cm}
\be
\cdot
\underbrace{g_{(a_1,b_1)}^{(i_1,j_1)}\,g_{(a_2,b_2)}^{(i_2,j_2)}\,g_{(a_3,b_3)}^{(i_3,j_3)}
\cdot K_{(a_1,b_1 )}^{(i_1,j_1 )}(N) K_{( a_2,b_2)}^{( i_2,j_2)}(N)K_{( a_3,b_3)}^{(
i_3,j_3)}(N)}
_{f_{(a_1,b_1)}^{(i_1,j_1)}\,f_{(a_2,b_2)}^{(i_2,j_2)}\,f_{(a_3,b_3)}^{(i_3,j_3)}}
\cdot\,\xi_{(a_1,b_1|a_2,b_2|a_3,b_3)}^{(i_1,j_1|i_2,j_2|i_3,j_3)}
\ee
In the first approximation the correction factors in the 3-hook case are
(they are {\bf never literally true}, before $\eta$-factors are introduced):
\be
\xi_{(a_1,b_1|a_2,b_2|a_3,b_3)}^{(i_1,j_1)} =
{\cal K}_{(a_1,b_1|a_2,b_2|a_3,b_3)}^{(i_1,j_1)}
 \big(\underline{N }\big)
\cdot\ \boxed{(1-\delta_{i_1\cdot j_1})}\  +
\nn\ee
\vspace{-0.5cm}
\be
+  \frac{[N+a_2-j_1][N-b_2+i_1]}{[N+a_2+i_1+1][N-b_2-j_1-1]}
\cdot
\frac{[N+a_3-j_1][N-b_3+i_1]}{[N+a_3+i_1+1][N-b_3-j_1-1]}
\cdot
{\cal K}_{(a_1,b_1|a_2,b_2|a_3,b_3)}^{(i_1,j_1)}(\underline{N+i_1-j_1})
\cdot \boxed{\delta_{i_1\cdot j_1}}\
\nn
\ee
\be
\xi_{(a_1,b_1|a_2,b_2|a_3,b_3)}^{(i_1,j_1|i_2,j_2)} =
{\cal K}_{(a_1,b_1|a_2,b_2|a_3,b_3)}^{(i_1,j_1|i_2,j_2)}
\big(\underline{N}\big)
\label{xi3prot}
\ee
(note that $a_2>0$ and $b_2>0$ for 3-hook diagrams thus the shifts like
$N\longrightarrow \underline{N+(i_1+1)\delta_{b_2}-(j_1+1)\delta_{a_2})}$
do not matter)
and
\be
\xi_{(a_1,b_1|a_2,b_2|a_3,b_3)}^{(i_1,j_1|i_2,j_2|i_3,j_3)} =
\nn
{\cal K}_{(a_1,b_1|a_2,b_2|a_3,b_3)}^{(i_1,j_1|i_2,j_2|i_3,j_3)}
\Big(\underline{N+(i_1+i_2+i_3+3)\cdot \delta_{b_3 } - (j_1+j_2+j_3+3)\cdot \delta_{a_3
}}\Big)
\ee
with
\vspace{-0.5cm}
\be
{\cal K}_{(a_1,b_1|a_2,b_2|a_3,b_3)}^{(i_1,j_1)}(\underline{N})=
\frac{{K}_{(a_1,b_1|a_2,b_2|a_3,b_3)}^{(i_1,j_1)}(\underline{N})}
{K_{(a_1,b_1 )}^{(i_1,j_1)}(\underline{N})}
\nn \\
{\cal K}_{(a_1,b_1|a_2,b_2|a_3,b_3)}^{(i_1,j_1|i_2,j_2)}(\underline{N})=
\frac{{K}_{(a_1,b_1|a_2,b_2|a_3,b_3)}^{(i_1,j_1|i_2,j_2)}(\underline{N})}
{K_{(a_1,b_1 )}^{(i_1,j_1)}(\underline{N})\cdot K_{(a_2,b_2 )}^{(i_2,j_2)}(\underline{N})}
\nn \\
{\cal K}_{(a_1,b_1|a_2,b_2|a_3,b_3)}^{(i_1,j_1|i_2,j_2|i_3,j_3)}(\underline{N})=
\frac{{K}_{(a_1,b_1|a_2,b_2|a_3,b_3)}^{(i_1,j_1|i_2,j_2|i_3,j_3)}(\underline{N})}
{K_{(a_1,b_1 )}^{(i_1,j_1)}(\underline{N})\cdot K_{(a_2,b_2
)}^{(i_2,j_2)}(\underline{N})\cdot
K_{(a_3,b_3 )}^{(i_3,j_3)}(\underline{N})}
\ee

Correction factors $\eta_{(22|11|00)}^\mu$ appear to be
\be
\eta^{(00)}=\eta^{(01)}=\eta^{(10)}=\eta^{(20)}=\eta^{(02)}=\eta^{(22)}=1 \nn \\
\eta^{(11)} = \frac{D_1D_0^2D_{-1}}{D_2^2D_{-2}^2} \ \ \ \ \ \ \
\eta^{(12)}=  \frac{D_0^2}{D_2D_{-2}} =
\frac{{\cal K}_{(22|11|00)}^{(12)}(\underline{N+2})}
{{\cal K}_{(22|11|00)}^{(12)}(\underline{N})} \ \ \ \ \
\eta^{(21)} = \frac{D_0^2}{D_2D_{-2}} =
\frac{{\cal K}_{(22|11|00)}^{(21)}(\underline{N-2})}
{{\cal K}_{(22|11|00)}^{(21)}(\underline{N})} \nn \\
\eta^{(11|00)} = \frac{D_0^4}{D_3D_1D_{-1}D_{-3}} \nn \\
\eta^{(12|00)} = \frac{D_2D_0^4}{D_3D_1^2D_{-1}D_{-2}}
= \frac{{\cal K}_{(22|11|00)}^{(12|00)}(\underline{N+1})}
{{\cal K}_{(22|11|00)}^{(12|00)}(\underline{N})}\cdot\frac{D_1^2}{D_3D_{-1}}
\nn \\
\eta^{(21|00)} = \frac{D_0^4D_{-2}}{D_2D_1D_{-1}^2D_{-3}}
= \frac{{\cal K}_{(22|11|00)}^{(21|00)}(\underline{N-1})}
{{\cal K}_{(22|11|00)}^{(21|00)}(\underline{N})}\cdot \frac{D_{-1}^2}{D_1D_{-3}}
\nn \\
\eta^{(12|01)} = \frac{D_0^3}{D_3D_{-1}D_{-2}}
=\frac{{\cal K}_{(22|11|00)}^{(12|01)}(\underline{N+1})}
{{\cal K}_{(22|11|00)}^{(12|01)}(\underline{N})}\cdot \frac{D_1^2}{D_3D_{-1}} \nn \\
\eta^{(21|10)} = \frac{D_0^3}{D_2D_1D_{-3}} =
\frac{{\cal K}_{(22|11|00)}^{(12|10)}(\underline{N-1})}
{{\cal K}_{(22|11|00)}^{(12|01)}(\underline{N})}\cdot\frac{D_{-1}^2}{D_1D_{-3}}
\nn\\
\eta^{(22|00)}=\eta^{(22|01)}=\eta^{(22|10)}=\frac{D_0^2}{D_1D_{-1}} \nn \\
\eta^{(22|11)}=\eta^{(22|11|00)}=1
\ee
and
the answer for the  $F$-function is
\vspace{-0.3cm}
\be
A^{-9} \cdot F_{(22|11|00)}^{(m)} = \nn \\
=\frac{1}{D_2D_1^2D_0^3D_{-1}^2D_{-2}}\,\Big(1 - \Lambda_{(22|11|00)}^m\Big) -
\frac{[3]^2}{D_3D_2D_1D_0^3D_{-1}D_{-2}D_{-3}}
\,\left(\Lambda_{(00)}^m - \Lambda_{(22|11)}^m\right) +
\nn \\
+\frac{\frac{[4][3]^2}{[2]}}{D_3D_2^2D_1D_0 D_{-1}^2D_{-2}D_{-4}}
\,\left(\Lambda_{(01)}^m - \Lambda_{(22|10)}^m
\right) +
\frac{\frac{[4][3]^2}{[2]}}{D_4D_2D_1^2D_0D_{-1}D_{-2}^2D_{-3}}
\,\left(\Lambda_{(10)}^m - \Lambda_{(22|01)}^m
\right) -
\nn \\
-\frac{\frac{[5][4]}{[2]}}{D_3D_2^2D_1^2D_0 D_{-2}D_{-3}D_{-4}}
\,\left(\Lambda_{(02)}^m - \Lambda_{(21|10)}^m
\right) -
\frac{\frac{[5][4]}{[2]}}{D_4D_3D_2 D_0 D_{-1}^2D_{-2}^2D_{-3}}
\,\left(\Lambda_{(20)}^m - \Lambda_{(12|01)}^m
\right) -
\nn \\
-\frac{[4]^2[2]^2}{D_4D_2^2D_0^3D_{-2}^2D_{-4}}
\,\left(\Lambda_{(11)}^m - \Lambda_{(22|00)}^m
\right) +
\nn \\
+\frac{[5][3]^2}{D_4D_2^2D_1D_0D_{-1}^2D_{-3}D_{-4}}
\,\left(\Lambda_{(12)}^m  - \Lambda_{(21|00)}^m
\right) +
\frac{[5][3]^2}{D_4D_3D_1^2D_0D_{-1}D_{-2}^2D_{-4}}
\left(\Lambda_{(21)}^m  - \Lambda_{(12|00)}^m
\right) -
\nn \\
-\frac{\frac{[4]^2[3]^2}{[2]^2}}{D_4D_3D_1D_0^3D_{-1}D_{-3}D_{-4}}\,
\left(\Lambda_{(22)}^m  - \Lambda_{(11|00)}^m
\right)
\ee
This is actually a Laurent polynomial at all $m$, satisfying
(\ref{sumrules1}).

\section{Extension to $F_{(a_1b_1|11|00)}$}

\noindent
Again, we can easily extend this result to arbitrary $a_1$ and $b_1$:
the substitute of (\ref{nonskew}),
$\ {\bf true\ for\ a_2\cdot b_2=1,\ a_3\cdot b_3=0}$, is
\vspace{-0.5cm}
\be
\xi_{(a_1,b_1|a_2,b_2|a_3,b_3)}^{(i_1,j_1)} =
\ {\cal K}_{(a_1,b_1|a_2,b_2|a_3,b_3)}^{(i_1,j_1)}
 \Big(\underline{N+2(\delta_{i_1-1}-\delta_{j_1-1}) }\Big)
\cdot\left(\frac{u}{K_{(a_1,b_1|a_2,b_2|a_3,b_3)}^{(i_1,j_1)}}\right)^{\delta_{i_1\cdot
j_1-1}}
\!\!\!\!\!\!\cdot\ \boxed{(1-\delta_{i_1\cdot j_1}) } \ +
\nn\ee
\vspace{-0.6cm}
\be
\!\!\!\!\!\!\!
+  \frac{[N+a_2-j_1][N-b_2+i_1]}{[N+a_2+i_1+1][N-b_2-j_1-1]}
\cdot
\frac{[N+a_3-j_1][N-b_3+i_1]}{[N+a_3+i_1+1][N-b_3-j_1-1]}
\cdot
{\cal K}_{(a_1,b_1|a_2,b_2|a_3,b_3)}^{(i_1,j_1)}(\underline{N+i_1-j_1})
\cdot \boxed{\delta_{i_1\cdot j_1}}
\nn\label{xi1b}
\ee
\vspace{-0.2cm}
\be
\!\!\!\!\!\!\!\!\!\!\!\!\!\!\!\!\!\!\!\!\!\!
\xi_{(a_1,b_1|a_2,b_2|a_3,b_3)}^{(i_1,j_1|i_2,j_2)} =
\label{xi2b}
{\cal K}_{(a_1,b_1|a_2,b_2|a_3,b_3)}^{(i_1,j_1|i_2,j_2)}
\Big(\underline{N+\delta_{i_1,1}-\delta_{j_1,1}}\Big)
\cdot \left(\frac{D_1^2}{D_3D_{-1}}\right)^{\delta_{i_1-1}}
\!\!\!\! \cdot\left(\frac{D_{-1}^2}{D_1D_3}\right)^{\delta_{j_1-1}}
\!\!\!\! \cdot \left(\frac{D_0^2}{D_1D_{-1}}\right)^{2\delta_{i_1\cdot j_1-1}
+ (1-\delta_{i_1-1})\cdot(1-\delta_{ j_1-1}))
\cdot\delta_{i_2\cdot j_2}}
\nn\ee
\vspace{-0.2cm}
\be
\xi_{(a_1,b_1|a_2,b_2|a_3,b_3)}^{(i_1,j_1|i_2,j_2|i_3,j_3)} =
\label{xi3b}
{\cal K}_{(a_1,b_1|a_2,b_2|a_3,b_3)}^{(i_1,j_1|i_2,j_2|i_3,j_3)}
\Big(\underline{N
+(i_1+i_2+i_3+3)\cdot \delta_{b_3} - (j_1+j_2+j_3+3)\cdot \delta_{a_3}
}\Big)
\ee

\noindent
The  shift $N \ \longrightarrow \ \underline{N
+(i_1+i_2+i_3+3)\cdot \delta_{b_3} - (j_1+j_2+j_3+3)\cdot \delta_{a_3}}$
in the last line is not actually tested by these formulas, because the
associated ${\cal K}_{(a_1b_1|11|00)}^{(i_1j_1|11|00)}$ do not depend on $A$.

The quantity $u_{(a_1b_1|11|00)}$ is given by a literal analogue of
(\ref{nonskew}):
\be
\boxed{
u_{(a_1,b_1|1,1|0,0)} = \left({\cal K}_{(a_1,b_1|1,1|0,0)}^{(1,1)}
- \frac{[a_1+2]\,[b_1+2]}{[a_1]\,[b_1]}
\cdot\frac{[3][2]^2\{q\}^2}{D_0^2}\right)\cdot
\frac{D_1D_0^2D_{-1}}{D_3D_2D_{-2}D_{-3}}
}
\label{nonskew3}
\ee

\section{Racah matrix $\bar S$ for representation $R=[333]$}

Coming back to the case of $R=[333]$ we can now use (\ref{HRdb1})
to get the matrix elements  $\bar S^{[333]}_{\mu\nu}$.
For this purpose it is technically convenient to substitute
the expansion in $\Lambda_\mu^m \Lambda_\nu^n$ by that in $\Lambda_\mu\bar\Lambda_\nu$
with independent $\bar \Lambda$ and $\lambda$ instead of arbitrary $m$ and $n$.
To get a $20\times 20$ matrix we need to enumerate the subdiagrams of $R=[333]$,
which are also in one-to-one correspondence with the $20$ irreducible representations
in $R\otimes \bar R= [333] \otimes \overline{[333]}$:
\be
\nn\\
\begin{array}{|c|ccc|cccccc|}
\hline
&&&&&&&&&\\
1 & 2 & 3 & 4 & 5 & 6 & 7 & 8 & 9 & 10 \\
&&&&&&&&&\\
\emptyset & [1] & [11] & [111] & [2] & [21] & [211]  &[22] & [221] & [222]  \\
&&&&&&&&&\\
\emptyset & (00) & (01) & (02) & (10) & (11) & (12) & (11|00) & (12|00) & (12|01)  \\
&&&&&&&&&\\
\hline
\end{array}
\nn\ee
\be
\begin{array}{|cccccccccc|}
\hline
&&&&&&&&&\\
11 & 12 & 13 & 14 & 15 & 16 & 17 & 18 & 19 & 20 \\
&&&&&&&&&\\
\phantom. [3] & [31] & [311] & [32] & [321] & [322]  &[33] & [331] & [332] & [333]  \\
&&&&&&&&&\\
(20) & (21) & (22) & (21|00) & (22|00) & (21|10) & (22|01) & (22|10) & (22|11) &
(22|11|00)\\
&&&&&&&&&\\
\hline
\end{array}
\nn \ee

Dimensions ${\cal D}_\mu$ of these representations are obtained from
the terms with $\nu=\emptyset$ in (\ref{HRdb1}), because
$\bar S_{\mu\emptyset} = \frac{\sqrt{{\cal D}_\mu}}{d_R}$:
in obvious notation
\be
{\cal D}_\mu = d_R^2 \cdot {\rm coeff}(H_R, \Lambda_\mu\bar \Lambda_\emptyset)
\ee
\vspace{-0.5cm}
After that
\be
\bar S_{\mu\nu} = \frac{d_R}{\sqrt{\cal D_\mu\,\cal D_\nu}}\cdot
\cdot {\rm coeff}(H_R, \Lambda_\mu\bar \Lambda_nu)
\ee
The simplest test of the result is that $\bar S$ is orthogonal matrix,
\vspace{-0.2cm}
\be
\sum_{\nu=1}^{20} \bar S_{\mu\nu} \bar S_{\mu'\nu} = \delta_{\mu\mu'}
\ee
It is also symmetric.

The second exclusive matrix $S^{[333]}$ is then  the diagonalizing matrix of
$\bar T\bar S\bar T$ \cite{arbor}:
\be
\bar T\bar S \bar T= S T^{-1} S^\dagger
\ee
with the known diagonal $T$ and $\bar T$, made from the $q$-powers of Casimir.
This is actually a linear   equation for $S$,
\be
\Big(\bar T\bar S \bar T\Big)\,S = S\, T^{-1}
\label{SfrobS}
\ee
which is practically solvable,
though explicit calculation is somewhat tedious.
The resulting matrix $S^{[333]}_{\mu\nu}$ should be orthogonal --
what fixes the normalization of the solution to (\ref{SfrobS}).
In variance with $\bar S$, this $S$ is not symmetric.

\bigskip

A typical example of the matrix element is
{\footnotesize
$$
\bar S^{[333]}_{7,15} = \bar S^{[333]}_{[211],[321]} =
-\frac{[5]\cdot \{q\}}{D_4D_2^2D_0D_{-2}D_{-4}}\cdot\sqrt{\frac{D_5D_3}{D_1D_{-1}}}\cdot
$$
$$\cdot \Big(A^6q^{-2}
-A^4(2q^8+3q^6+2q^4+q^2-3-5q^{-2}-2q^{-4}+2q^{-6}+2q^{-8}+3q^{-10}+q^{-12})
+
$$
$$
+A^2\left(q^{18}+3q^{16}+4q^{14}+4q^{12}-6q^8-9q^6-5q^4+2q^2+12
+{13}q^{-2}+{5}q^{-4}-{4}q^{-6}-{9}{q^{-8}}
-7q^{-10}-{q^{-12}}+{4}{q^{-14}}+{4}{q^{-16}}
+{3}{q^{-18}}+{q^{-20}}\right)
-
$$
{\tiny
$$
-(q^2+q^{-2})(q^{22}+3q^{20}+3q^{18}-q^{16}-8q^{14}-9q^{12}+q^{10}+14q^8+19q^6+6q^4-13q^2-22
-13q^{-2}+6q^{-4}+19q^{-6}+14q^{-8}+q^{-10}-9q^{-12}-8q^{-14}-q^{-16}+3q^{-18}+3q^{-20}+q^{-22})
-
$$
}
\vspace{-0.3cm}
$$
+A^{-2}(q^{-18}+3q^{-16}+4q^{-14}+4q^{-12}-6q^{-8}-9q^{-6}-5q^{-4}+2q^{-2}
+12+13q^2+5q^4-4q^6-9q^8
-7q^{10}-q^{12}+4q^{14}+4q^{16}+3q^{18}+q^{20})
-
$$
$$
-A^{-4}(2q^{-8}+3q^{-6}+2q^{-4}+q^{-2}-3-5q^2-2q^4+2q^6+2q^8+3q^{10}+q^{12})
+A^{-6}q^2\Big)
$$
}
The polynomial in brackets reduces to $D(0)^6=\{A\}^6$ at $q=1$ and to
$\ -\left([4][3][2]\right)^3\{q\}^6\ $ at $A=1$.
A better quantity for practical calculations is unnormalized
$\bar \sigma_{\mu\nu} =  \bar S_{\mu\nu}\cdot \sqrt{{\cal D}_\mu{\cal D}_\nu} $,
which does not contain square roots.

\section{Conclusion
\label{conc}}

The main result of the present letter is explicit expression for the
two previously unknown $F$-functions $F_{(22|11)}^{(m)}$ and $F_{(22|11|00)}^{(m)}$.
Most important is the deviation from the coefficient $f_{(22|11)}^{(11)}$
from the skew dimension, even shifted -- what is expressed by eq.(\ref{nonskew}),
see also (\ref{nonskew3}).
This new phenomenon explains the failure of previous naive attempts
to write down an explicit general expression
for $F$ in arbitrary representation:
an adequate substitute of the skew characters and appropriate generalization of
the corresponding conjecture in \cite{KMtwist} is needed for this.
The next step in this study should be further extension to $a_2\cdot b_2>1$.

The two newly-found functions, if combined with the other $18$,
associated with 0,1,2-hook diagrams $\lambda$ with the property $a_2\cdot b_2=0$,
provide explicit expression for $[333]$-colored HOMFLY for all twist and double braid
knots.
Moreover, from (\ref{HRdb1}) one can read all the elements of the Racah matrix
$\bar S^{[333]}$, while $S^{[333]}$ is then found from (\ref{SfrobS}).
Thus this paper solves the long-standing problem to evaluate
$\bar S^{[333]}$ and $S^{[333]}$.
Explicit expressions for these Racah matrices as well as for the  $[333]$-colored HOMFLY
for the simplest twist and double-braid knots are available at \cite{knotebook}.

It still remains to evaluate the twist-knot polynomials and Racah matrices for {\it generic}
rectangular representations -- the new step, made in the present paper, provides the
essential new knowledge about this problem which can help to overcome the
existing deadlock.\footnote{ Comment to version 3: This problem is now solved in \cite{KNTZ}
and \cite{RectSbar}.}

For additional peculiarities of {\it non-rectangular} case see \cite{Mnonrect}.
The main point there is that representations in $R\otimes \bar R$ are no longer
in one-to-one correspondence with the sub-diagrams of non-rectangular $R$.
Still, factorization of the coefficients in the differential expansion
for double braids persists, and thus the Racah matrices $\bar S$ can still
be extracted from knot polynomials -- though the procedure becomes more tedious
\cite{MnonrectS}.

\section*{Acknowledgements}

This work was performed at the Institute for the Information Transmission Problems
with the support from the Russian Science Foundation, Grant No.14-50-00150.

\end{document}